\documentclass[a4paper,11pt]{article}
\usepackage[T1]{fontenc} % if needed
\usepackage{float} % if needed

\makeatletter
\gdef\@fpheader{ }
\gdef\@journal{ }
\makeatother
\RequirePackage{amsmath}
\RequirePackage{amssymb}
\RequirePackage{epsfig}
\RequirePackage{graphicx}
\RequirePackage[numbers,sort&compress]{natbib}
\RequirePackage{color}
\RequirePackage[colorlinks=true
,urlcolor=blue
,anchorcolor=blue
,citecolor=blue
,filecolor=blue
,linkcolor=blue
,menucolor=blue
,pagecolor=blue
,linktocpage=true
,pdfproducer=medialab
,pdfa=true
]{hyperref}

\newif\ifnotoc\notocfalse
\newif\ifemailadd\emailaddfalse
\newif\iftoccontinuous\toccontinuousfalse
\makeatletter
\def\@subheader{\@empty}
\def\@keywords{\@empty}
\def\@abstract{\@empty}
\def\@xtum{\@empty}
\def\@dedicated{\@empty}
\def\@arxivnumber{\@empty}
\def\@collaboration{\@empty}
\def\@collaborationImg{\@empty}
\def\@proceeding{\@empty}
\def\@preprint{\@empty}

\newcommand{\subheader}[1]{\gdef\@subheader{#1}}
\newcommand{\keywords}[1]{\if!\@keywords!\gdef\@keywords{#1}\else%
\PackageWarningNoLine{\jname}{Keywords already defined.\MessageBreak Ignoring last definition.}\fi}
\renewcommand{\abstract}[1]{\gdef\@abstract{#1}}
\newcommand{\dedicated}[1]{\gdef\@dedicated{#1}}
\newcommand{\arxivnumber}[1]{\gdef\@arxivnumber{#1}}
\newcommand{\proceeding}[1]{\gdef\@proceeding{#1}}
\newcommand{\xtumfont}[1]{\textsc{#1}}
\newcommand{\correctionref}[3]{\gdef\@xtum{\xtumfont{#1} \href{#2}{#3}}}
\newcommand\jname{JHEP}

\newcommand\preprint[1]{\gdef\@preprint{\hfill #1}}

\makeatother

\newcommand\note[2][]{%
\if!#1!%
\stepcounter{footnote}\footnotetext{#2}%
\else%
{\renewcommand\thefootnote{#1}%
\footnotetext{#2}}%
\fi}

\makeatletter
\newtoks\auth@toks
\renewcommand{\author}[2][]{%
  \if!#1!%
    \auth@toks=\expandafter{\the\auth@toks#2\ }%
  \else
    \auth@toks=\expandafter{\the\auth@toks#2$^{#1}$\ }%
  \fi
}
\makeatother
\makeatletter
\newtoks\affil@toks\newif\ifaffil\affilfalse
\newcommand{\affiliation}[2][]{%
\affiltrue
  \if!#1!%
    \affil@toks=\expandafter{\the\affil@toks{\item[]#2}}%
  \else
    \affil@toks=\expandafter{\the\affil@toks{\item[$^{#1}$]#2}}%
  \fi
}
\makeatother
\makeatletter
\newtoks\email@toks\newcounter{email@counter}%
\newcommand{\emailAdd}[1]{%
\emailaddtrue%
\ifnum\theemail@counter>0\email@toks=\expandafter{\the\email@toks, \@email{#1}}%
\else\email@toks=\expandafter{\the\email@toks\@email{#1}}%
\fi\stepcounter{email@counter}}
\newcommand{\@email}[1]{\href{mailto:#1}{\tt #1}}
\makeatother

\makeatletter
\newcommand*\collaboration[1]{\gdef\@collaboration{#1}}
\newcommand*\collaborationImg[2][]{\gdef\@collaborationImg{#2}}
\makeatletter
\newcommand\afterLogoSpace{\smallskip}
\newcommand\afterSubheaderSpace{\vskip3pt plus 2pt minus 1pt}
\newcommand\afterProceedingsSpace{\vskip21pt plus0.4fil minus15pt}
\newcommand\afterTitleSpace{\vskip23pt plus0.06fil minus13pt}
\newcommand\afterRuleSpace{\vskip23pt plus0.06fil minus13pt}
\newcommand\afterCollaborationSpace{\vskip3pt plus 2pt minus 1pt}
\newcommand\afterCollaborationImgSpace{\vskip3pt plus 2pt minus 1pt}
\newcommand\afterAuthorSpace{\vskip5pt plus4pt minus4pt}
\newcommand\afterAffiliationSpace{\vskip3pt plus3pt}
\newcommand\afterEmailSpace{\vskip16pt plus9pt minus10pt\filbreak}
\newcommand\afterXtumSpace{\par\bigskip}
\newcommand\afterAbstractSpace{\vskip16pt plus9pt minus13pt}
\newcommand\afterKeywordsSpace{\vskip16pt plus9pt minus13pt}
\newcommand\afterArxivSpace{\vskip3pt plus0.01fil minus10pt}
\newcommand\afterDedicatedSpace{\vskip0pt plus0.01fil}
\newcommand\afterTocSpace{\bigskip\medskip}
\newcommand\afterTocRuleSpace{\bigskip\bigskip}
\newlength{\affiliationsSep}
\newcommand\beforetochook{\pagestyle{myplain}\pagenumbering{roman}}

\DeclareFixedFont\trfont{OT1}{phv}{b}{sc}{11}

\renewcommand\maketitle{
\pagestyle{empty}
\thispagestyle{titlepage}
\setcounter{page}{0}
\noindent{\small\scshape\@fpheader}\@preprint\par

\afterLogoSpace
\if!\@subheader!\else\noindent{\trfont{\@subheader}}\fi
\afterSubheaderSpace
\if!\@proceeding!\else\noindent{\sc\@proceeding}\fi
\afterProceedingsSpace
{\LARGE\flushleft\sffamily\bfseries\@title\par}
\afterTitleSpace
\hrule height 1.5\p@%
\afterRuleSpace
\if!\@collaboration!\else
{\Large\bfseries\sffamily\raggedright\@collaboration}\par
\afterCollaborationSpace
\fi
\if!\@collaborationImg!\else
{\normalsize\bfseries\sffamily\raggedright\@collaborationImg}\par
\afterCollaborationImgSpace
\fi
{\bfseries\raggedright\sffamily\the\auth@toks\par}
\afterAuthorSpace
\ifaffil\begin{list}{}{%
\setlength{\leftmargin}{0.28cm}%
\setlength{\labelsep}{0pt}%
\setlength{\itemsep}{\affiliationsSep}%
\setlength{\topsep}{-\parskip}}
\itshape\small%
\the\affil@toks
\end{list}\fi
\afterAffiliationSpace
\ifemailadd %% if emailadd is true
\noindent\hspace{0.28cm}\begin{minipage}[l]{.9\textwidth}
\begin{flushleft}
\textit{E-mail:} \the\email@toks
\end{flushleft}
\end{minipage}
\else %% if emailaddfalse do nothing
\PackageWarningNoLine{\jname}{E-mails are missing.\MessageBreak Plese use \protect\emailAdd\space macro to provide e-mails.}
\fi
\afterEmailSpace
\if!\@xtum!\else\noindent{\@xtum}\afterXtumSpace\fi
\if!\@abstract!\else\noindent{\renewcommand\baselinestretch{.9}\textsc{Abstract:}}\ \@abstract\afterAbstractSpace\fi
\if!\@keywords!\else\noindent{\textsc{Keywords:}} \@keywords\afterKeywordsSpace\fi
\if!\@arxivnumber!\else\noindent{\textsc{ArXiv ePrint:}} \href{http://arxiv.org/abs/\@arxivnumber}{\@arxivnumber}\afterArxivSpace\fi
\if!\@dedicated!\else\vbox{\small\it\raggedleft\@dedicated}\afterDedicatedSpace\fi
\ifnotoc\else
\iftoccontinuous\else\newpage\fi
\beforetochook\hrule
\tableofcontents
\afterTocSpace
\hrule
\afterTocRuleSpace
\fi
\setcounter{footnote}{0}
\pagestyle{myplain}\pagenumbering{arabic}
} % close the \renewcommand\maketitle{

\renewcommand{\baselinestretch}{1.1}\normalsize
\divide\@tempdima\baselineskip
\@tempcnta=\@tempdima
\skip\@mpfootins = \skip\footins
\renewcommand{\@dotsep}{10000}

\newcommand\ps@myplain{
\pagenumbering{arabic}
\renewcommand\@oddfoot{\hfill-- \thepage\ --\hfill}
\renewcommand\@oddhead{}}
\let\ps@plain=\ps@myplain

\newcommand\ps@titlepage{\renewcommand\@oddfoot{}\renewcommand\@oddhead{}}

\numberwithin{equation}{section}

\renewcommand\section{\@startsection{section}{1}{\z@}%
                                   {-3.5ex \@plus -1.3ex \@minus -.7ex}%
                                   {2.3ex \@plus.4ex \@minus .4ex}%
                                   {\normalfont\large\bfseries}}
\renewcommand\subsection{\@startsection{subsection}{2}{\z@}%
                                   {-2.3ex\@plus -1ex \@minus -.5ex}%
                                   {1.2ex \@plus .3ex \@minus .3ex}%
                                   {\normalfont\normalsize\bfseries}}
\renewcommand\subsubsection{\@startsection{subsubsection}{3}{\z@}%
                                   {-2.3ex\@plus -1ex \@minus -.5ex}%
                                   {1ex \@plus .2ex \@minus .2ex}%
                                   {\normalfont\normalsize\bfseries}}
\renewcommand\paragraph{\@startsection{paragraph}{4}{\z@}%
                                   {1.75ex \@plus1ex \@minus.2ex}%
                                   {-1em}%
                                   {\normalfont\normalsize\bfseries}}
\renewcommand\subparagraph{\@startsection{subparagraph}{5}{\parindent}%
                                   {1.75ex \@plus1ex \@minus .2ex}%
                                   {-1em}%
                                   {\normalfont\normalsize\bfseries}}

\def\fnum@figure{\textbf{\figurename\nobreakspace\thefigure}}
\def\fnum@table{\textbf{\tablename\nobreakspace\thetable}}

\long\def\@makecaption#1#2{%
  \vskip\abovecaptionskip
  \sbox\@tempboxa{\small #1. #2}%
  \ifdim \wd\@tempboxa >\hsize
    \small #1. #2\par
  \else
    \global \@minipagefalse
    \hb@xt@\hsize{\hfil\box\@tempboxa\hfil}%
  \fi
  \vskip\belowcaptionskip}

\renewenvironment{thebibliography}[1]{%
\begin{oldthebibliography}{#1}%
\small%
\raggedright%
\setlength{\itemsep}{5pt plus 0.2ex minus 0.05ex}%
}%
{%
\end{oldthebibliography}%
}

\begin{document}

%%%%%%%%%%%%%%%%%%炎籾匈%%%%%%%%%%%%%%%%%%%%%%%%%%%%%&&&&&&&&&&&&&&&&&&&&&&

\title{\boldmath A group method solving many-body systems in intermediate statistical
representation }

% more complex case: 4 authors, 3 institutions, 2 footnotes

\author[a,1]{Yao Shen}\note{shenyaophysics@hotmail.com}
\author[b,2,*]{Chi-Chun Zhou}\note{zhouchichun@dali.edu.cn (corresponding author)}
\author[c,3]{Wu-sheng Dai}\note{daiwusheng@tju.edu.cn.}
\author[c,4]{and Mi Xie}\note{xiemi@tju.edu.cn.}
% The "\note" macro will give a warning: "Ignoring empty anchor..."
% you can safely ignore it.

\affiliation[a]{School of Criminal Investigation, People's Public Security University
of China, Beijing 100038, PR China}
\affiliation[b]{School of Engineering, Dali University, Dali, Yunnan 671003, PR China}
\affiliation[c]{Department of Physics, Tianjin University, Tianjin 300350, PR China}

%\affiliation[c]{DP School}

% e-mail addresses: one for each author, in the same order as the authors
%\emailAdd{Ccc@one.edu.cn}
%\emailAdd{second@asas.edu}
%\emailAdd{daiwusheng@tju.edu.cn}
%\emailAdd{fourth@one.univ}

%\title{\boldmath A title with some math: $x=1$}
%% %simple case: 2 authors, same institution
%% \author{A. Uthor}
%% \author{and A. Nother Author}
%% \affiliation{Institution,\\Address, Country}

% more complex case: 4 authors, 3 institutions, 2 footnotes
%\author[a,b,1]{F. Irst,\note{Corresponding author.}}
%\author[c]{S. Econd,}
%\author[a,2]{T. Hird\note{Also at Some University.}}
%\author[a,2]{and Fourth}

% The "\note" macro will give a warning: "Ignoring empty anchor..."
% you can safely ignore it.

%\affiliation[a]{One University,\\some-street, Country}
%\affiliation[b]{Another University,\\different-address, Country}
%\affiliation[c]{A School for Advanced Studies,\\some-location, Country}

% e-mail addresses: one for each author, in the same order as the authors

%\emailAdd{first@one.univ}
%\emailAdd{second@asas.edu}
%\emailAdd{third@one.univ}
%\emailAdd{fourth@one.univ}

%\date{date}

\abstract{The exact solution of the interacting many-body system is important
and is difficult to solve. In this paper, we introduce a group method
to solve the interacting many-body problem using the relation between
the permutation group and the unitary group. We prove a group theorem
first, then using the theorem, we represent the Hamiltonian of the
interacting many-body system by the Casimir operators of unitary group.
The eigenvalues of Casimir operators could give the exact values of
energy and thus solve those problems exactly. This method maps the
interacting many-body system onto an intermediate statistical representation.
We give the relation between the conjugacy-class operator of permutation
group and the Casimir operator of unitary group in the intermediate
statistical representation, called the Gentile representation. Bose
and Fermi cases are two limitations of the Gentile representation.
We also discuss the representation space of symmetric and unitary
group in the Gentile representation and give an example of the Heisenberg
model to demonstrate this method. It is shown that this method is
effective to solve interacting many-body problems.}

\maketitle
\flushbottom
%%%%%%%%%%%%%%%%%%炎籾匈潤崩%%%%%%%%%%%%%%%%%%%%%%%%%%%%%&&&&&&&&&&&&&&&&&&&

%%%%%%%%%%屎猟蝕兵

\section{Introduction}

To give the exact solution of the interacting many-body problem is
always difficult in physics. Few low dimensional systems could be
solved exactly, such as one dimensional Ising model \cite{aa0,aa}.
Usually, some simplified methods or one and two transformations are
used to get the approximate solution, for example the mean field method
which considers the far particle-particle interaction as a field effect,
and the Jordan-Wigner transformation which maps the interaction of
particles onto a fermion representation \cite{aa0,aa}. Intermediate
statistics is a hot topic in physics these decades, e.g., in topological
quantum computation \cite{x11,x12,x13,s1,s2}, quantum material \cite{key-7,key-8},
and quantum information\cite{q}. If the interacting many-body system
is intermediate statistical system, getting the approximate solution
or even the low dimensional exact solution becomes more difficult.
Gentile statistics is a kind of intermediate statistics which is named
after Gentile \cite{x5}. The maximum occupation number of Gentile
statistics is a finite number $n$\cite{x5,x6,x7}. In Gentile statistics,
the states are labelled by the practical occupation number, which
makes it easy to research. It can be used as an effective tool to
solve the interacting many-body problems. Gentile statistics goes
back to Bose and Fermi statistics in limit cases. The Gentile system
is not just a theoretical model, more and more real systems are found
obeying Gentile statistics, such as conjugate annulenes \cite{m}
and spin waves of magnetic systems \cite{m1,m2,m3,sp,sp2}.

We deal with the intermediate statistical system or normal many-body
spin system in Gentile representation. It is a Gentile realization.
For instance, the many-body system is an anyon system. Anyon statistics
could give the wave function an additional phase by braiding two different
types of anyons \cite{b,a,a1}. This phase depends on the winding
number and the statistical parameter. The system of anyons is always
complicated. In this case, we can convert the problem of the winding
number representation \cite{me} of anyon into the occupation number
representation of Gentile statistics. In the occupation number representation
of Gentile statistics, the many-body problem is easier to solve in
certain cases. At the same time, in this paper, we give a Gentile
realization of the Heisenberg spin model. This model describes the
long-range interaction.

Group method is useful to deal with many-body interacting systems
\cite{c,c1}. In this paper, we construct the relation between the
permutation group and the unitary group in the Gentile representation.
According to the relation between the Casimir operator of the unitary
group and the conjugacy-class operator of the permutation group in
Gentile statistics presented in the paper, it is easy to solve the
Hamiltonian of the many-body interaction system.

This paper is organized as follows. In section $2$, we introduce
the relation between permutation group and unitary group in the Gentile
representation. In section $3$, the representation space of permutation
group and unitary group in the Gentile representation is discussed.
In section $4$, we solve the Heisenberg model in Gentile representation
as an example. Finally, in section $5$, we conclude our result and
make some discussions.

\section{The relation between permutation group and unitary group in Gentile
representation}

Any finite group is isomorphic to a certain subgroup of the permutation
group, any compact Lie group is isomorphic to a certain subgroup of
the unitary group, and all infinite groups are isomorphic to the general
linear groups. The relation between the permutation group and the unitary
group is important and useful. The conjugacy-class operator consists
of the generators of the permutation group and commutes with all group
elements. Also, the Casimir operator consists of the generators of
the unitary group and commutes with each group element\cite{c,c1}.
In order to rewrite the Hamiltonian of an interacting many-body system,
taking the Heisenberg model as an example, using the Casimir operator,
we prove a relation between the permutation group and the unitary
group. Using the theorem, we can solve exact solutions of the interacting
many-body problem.

The conjugacy-class operator is the sum of group elements over a conjugate
class\cite{c,c1}. $\left(a\right)=\left(a_{1},\cdots,a_{m}\right)$
with $a_{1}\geq\cdots\geq a_{m}\geq0$ denotes an integer partition
of integer $M$ and $a_{i}$ is the element of this integer partition\cite{z,z1,z2}.
The length of $\left(a\right)$ is $l_{\left(a\right)}=m$. For example,
$\left(a\right)=\left(4\right)$. The partitions for $\left(4\right)$
are $\left(4\right)$, $\left(3,1\right)$, $\left(2,2\right)$, $\left(2,1^{2}\right)$,
$\left(1^{4}\right)$. Each integer partition of the permutation group
gives a conjugacy-class operator \cite{z,z1,z2}. The conjugacy-class
operator $P(2,1^{\nu-2})$ denotes the exchange of any two particles
when the spin system includes $\nu$ particles, and $(2,1^{\nu-2})$
is the integer partition. For example the system consists of $\nu$
particles $\left\{ 1,2,\cdots i\cdots j\cdots,\nu\right\} $. Only
the $i$th and the $j$th particles are exchanged.

The operators which commute with all operators of certain group are
called the Casimir operators of this group \cite{c}. The number of
the Casimir operators equals to the rank of the group. The eigenvalues
of the Casimir operators represent the irreducible representations
\cite{c,c1}. The unitary group $U(m)$ has $m$ linear independent
Casimir operators. The Casimir operator of order $p$ reads \cite{c,c1}
\begin{equation}
C_{p}=\sum_{k_{1}\cdots k_{p}=1}^{m}E_{k_{1}k_{2}}\cdots E_{k_{p}k_{1}},
\end{equation}
where $E_{kl}$ is the generator of the unitary group $U(m)$. The
eigenvalue of the Casimir operator of order $p$ can be expressed
as
\begin{equation}
S_{p}=\sum_{i=1}^{m}\left[(a_{i}+m-i)^{p}-(m-i)^{p}\right],
\end{equation}
where $a_{i}$ are $m$ nonnegative integers that represent the irreducible
representation of $U(m)$ with $a_{1}\geq\cdots\geq a_{m}\geq0$ \cite{c,z,z1,z2}.

\textbf{Theorem:} The conjugacy-class operator $P(2,1^{\nu-2})$ of
permutation group $S_{N}$ and the Casimir operators $C_{1}$ and
$C_{2}$ (order one and two) of the unitary group $U(m)$ satisfy
\begin{equation}
Re[e^{\frac{i2\pi}{n+1}}P(2,1^{\nu-2})]+m\sum_{k=1}^{m}\sum_{i=1}^{\nu}J(N_{k}^{i})=\frac{1}{2}C_{2}-\frac{m}{2}C_{1},\label{eq:=000026}
\end{equation}
where 
\begin{equation}
J(N_{k}^{i})=-2\csc^{2}(\frac{\pi}{n+1})\sin\frac{\pi}{2(n+1)}\sin\frac{N_{k}^{i}\pi}{n+1}\sin\frac{(2N_{k}^{i}+n)\pi}{2(n+1)},\label{eq:=000026=000026}
\end{equation}
$n$ is the maximum occupation number of intermediate statistics,
$\nu$ is the practical particle number of one quantum state, and
$N_{k}^{i}$ is the particle number operator of the representation
(or energy) $k$ and particle (or position) $i$. When $n\rightarrow\infty$
(the Bose case) and $n=1$ (the Fermi case), $J(N_{k}^{i})=-N_{k}^{i}$. In
these two limitations, the relation becomes
\begin{equation}
\pm P(2,1^{\nu-2})-m\sum_{k=1}^{m}\sum_{i=1}^{\nu}N_{k}^{i}=\frac{1}{2}C_{2}-\frac{m}{2}C_{1},
\end{equation}
where $\pm$ correspond to bosons and fermions, respectively.

\textbf{Proof. }For Gentile statistics, we use two sets of creation
and annihilation operators $a^{\dagger},a$, $b^{\dagger},b$ and
$a=b^{*}$. This is because one set of creation and annihilation operators
is not enough to realize Gentile statistics. Let $a_{k}^{\dagger i}$
with superscript ranging from $1$ to $\nu$ and subscript ranging
from $1$ to $m$ represent creating the $i$th particle of state
$k$ (or creating a particle of energy $k$ at position $i$). The
relation of creation and annihilation operators follows the n-bracket
\cite{x6,x7}
\begin{equation}
\left[b_{k}^{i},a_{l}^{\dagger j}\right]_{n}=b_{k}^{i}a_{l}^{\dagger j}-e^{\frac{i2\pi}{n+1}}a_{l}^{\dagger j}b_{k}^{i}=\delta_{ij}\delta_{kl},
\end{equation}
where $\delta_{ij}$ is Kronecker symbol. 

At the same time, we have the operator relations
\begin{equation}
e^{i\pi/\left(  n+1\right)  }ba=ab,
\end{equation}
\begin{equation}
e^{i2\pi/\left(  n+1\right)  }b^{\dagger}a^{\dagger}=a^{\dagger}b^{\dagger}.
\end{equation}

The generator $\tau_{ij}$ (exchanging $i$th and $j$th particles)
of permutation group $S_{N}$ can be expressed in terms of creation
and annihilation operators as 
\begin{equation}
\tau_{ij}=\sum_{k,l=1}^{m}(a_{k}^{\dagger i}a_{l}^{\dagger j}b_{l}^{i}b_{k}^{j}+a_{k}^{\dagger i}b_{l}^{\dagger j}b_{l}^{i}a_{k}^{j}).\label{eq:a}
\end{equation}
The form of $\tau_{ij}$ is an obvious exchange of two particles (please
see next section). What calls for special attention is that the exchange
of two particles happens both in the Hilbert space and its conjugate
space. We will discuss in details in next section. So the conjugacy-class
operator $P(1^{\nu},2)$ of permutation group$S_{N}$ has the form
\begin{equation}
P(2,1^{\nu-2})=\sum_{i<j=1}^{\nu}\tau_{ij}=\sum_{i<j=1}^{\nu}\sum_{k,l=1}^{m}(a_{k}^{\dagger i}a_{l}^{\dagger j}b_{l}^{i}b_{k}^{j}+a_{k}^{\dagger i}b_{l}^{\dagger j}b_{l}^{i}a_{k}^{j}).
\end{equation}

We also construct the generator of unitary group $U(m)$ as
\begin{equation}
E_{kl}=\sum_{i=1}^{\nu}(a_{k}^{\dagger i}b_{l}^{i}+b_{k}^{\dagger i}a_{l}^{i}).\label{eq:e}
\end{equation}
In this case, the generator commutation relation of the unitary group
$U(m)$ satisfy
\begin{equation}
[E_{kl},E_{pq}]=\delta_{lp}E_{kq}-\delta_{qk}E_{pl}+2\delta_{lp}\delta_{qk}\underset{i}{\sum}[f(N_{l}^{i})g(N_{k}^{i})-f(N_{k}^{i})g(N_{l}^{i})],\label{eq:r}
\end{equation}
where
\begin{equation}
f(N_{k}^{i})=(aa^{\dagger}-a^{\dagger}a)_k^i=\csc\frac{\pi}{n+1}(\cos\frac{\pi}{n+1}-1)\sin\frac{N_{k}^{i}\pi}{n+1}+\cos\frac{N_{k}^{i}\pi}{n+1}\label{eq:r1}
\end{equation}
and
\begin{equation}
g(N_{k}^{i})=(a^{\dagger}a)_k^i=\csc\frac{\pi}{n+1}\sin\frac{N_{k}^{i}\pi}{n+1}.\label{eq:r2}
\end{equation}
For bosons $n\rightarrow\infty$ and fermions $n=1$, we have $g(N)=N$, $f(N)=-1$. The last two parts of Eq. (\ref{eq:r}) can cancel each other.
Then Eq. (\ref{eq:r}) turns to a familiar one
\begin{equation}
[E_{kl},E_{pq}]=\delta_{lp}E_{kq}-\delta_{qk}E_{pl}.\label{r3}
\end{equation}
Eq. (\ref{r3}) verifies that the construction equation (\ref{eq:e})
in Gentile statistics represents the Lie algebra of unitary group
$U(m)$.

The Casimir operator always relates to some invariances. This generator
insures that the Casimir operator of the unitary group $U(m)$ is
Hermit. In this case, we have the Casimir operators of order one
and two 
\begin{equation}
C_{2}=\sum_{k,l=1}^{m}E_{kl}E_{lk}=\sum_{k,l=1}^{m}\left[\sum_{i=1}^{\nu}(a_{k}^{\dagger i}b_{l}^{i}+b_{k}^{\dagger i}a_{l}^{i})\right]\left[\sum_{j=1}^{\nu}(a_{l}^{\dagger j}b_{k}^{j}+b_{l}^{\dagger j}a_{k}^{j})\right],
\end{equation}
\begin{equation}
C_{1}=\sum_{l=1}^{m}E_{ll}=\sum_{l=1}^{m}\sum_{i=1}^{\nu}(a_{l}^{\dagger i}b_{l}^{i}+b_{l}^{\dagger i}a_{l}^{i}).
\end{equation}
The creation and annihilation operators in Gentile statistics (intermediate
statistics) satisfy
\begin{equation}
\begin{cases}
\begin{array}{c}
a^{\dagger}\left|\nu\right\rangle _{n}=\sqrt{\left\langle \nu+1\right\rangle _{n}}\left|\nu+1\right\rangle _{n},\\
b^{\dagger}\left|\nu\right\rangle _{n}=\sqrt{\left\langle \nu+1\right\rangle _{n}^{*}}\left|\nu+1\right\rangle _{n},
\end{array} & \begin{array}{c}
b\left|\nu\right\rangle _{n}=\sqrt{\left\langle \nu\right\rangle _{n}}\left|\nu-1\right\rangle _{n},\\
a\left|\nu\right\rangle _{n}=\sqrt{\left\langle \nu\right\rangle _{n}^{*}}\left|\nu-1\right\rangle _{n},
\end{array}\end{cases}
\end{equation}
where 
\begin{equation}
\left\langle \nu\right\rangle _{n}=\frac{1-e^{\frac{i2\pi\nu}{n+1}}}{1-e^{\frac{i2\pi}{n+1}}}
\end{equation}
and $N$ is the particle number operator 
\begin{equation}
N\left|\nu\right\rangle _{n}=\nu\left|\nu\right\rangle _{n}.\label{eq:b}
\end{equation}
We also have 
\begin{equation}
\left[b_{k}^{i},b_{l}^{\dagger j}\right]_{n}=\left[a_{k}^{i},a_{l}^{\dagger j}\right]_{n}=f(N_{k}^{i})\delta_{ij}\delta_{kl}
\end{equation}
and 
\begin{equation}
[a^{\dagger}b^{\dagger}a^{\dagger}b^{\dagger},b^{\dagger}a^{\dagger}b^{\dagger}a^{\dagger}]_n=[abab,baba]_n=0.
\end{equation}
According to Eqs. (\ref{eq:a}-\ref{eq:b}), we prove Eqs. (\ref{eq:=000026})
and (\ref{eq:=000026=000026}). 

\section{The representation space of permutation group and unitary group}

To seek the exact solution, we discuss the representation space of
the interacting many-body system. The structure of the group space
is realized in the language of creation and annihilation operators.
The representation space of the permutation group and the unitary
group is usually expressed as the direct product of permutation and
unitary group $S_{N}\otimes U(m)$, which is called dual structure
\cite{s,c,c1}. In Gentile representation, we also have this result
with a little difference. Because according to equations (\ref{eq:a})
and (\ref{eq:e}), we have the relation of generators
\begin{equation}
[\tau_{ij},E_{st}]=0.
\end{equation}

\textbf{Proof.
\begin{equation}
\begin{array}{ccc}
\tau_{ij}E_{st} & = & \underset{kl}{\sum}(a_{k}^{\dagger i}a_{l}^{\dagger j}b_{l}^{i}b_{k}^{j}+a_{k}^{\dagger i}b_{l}^{\dagger j}b_{l}^{i}a_{k}^{j})\underset{r}{\sum}(a_{s}^{\dagger r}b_{t}^{r}+b_{s}^{\dagger r}a_{t}^{r})\\
 & = & \underset{klr}{\sum}(a_{k}^{\dagger i}a_{l}^{\dagger j}b_{l}^{i}b_{k}^{j}a_{s}^{\dagger r}b_{t}^{r}+a_{k}^{\dagger i}a_{l}^{\dagger j}b_{l}^{i}b_{k}^{j}b_{s}^{\dagger r}a_{t}^{r}+a_{k}^{\dagger i}b_{l}^{\dagger j}b_{l}^{i}a_{k}^{j}a_{s}^{\dagger r}b_{t}^{r}+a_{k}^{\dagger i}b_{l}^{\dagger j}b_{l}^{i}a_{k}^{j}b_{s}^{\dagger r}a_{t}^{r}),
\end{array}\label{eq:t1}
\end{equation}
\begin{equation}
\begin{array}{ccc}
E_{st}\tau_{ij} & = & \underset{r}{\sum}(a_{s}^{\dagger r}b_{t}^{r}+b_{s}^{\dagger r}a_{t}^{r})\underset{kl}{\sum}(a_{k}^{\dagger i}a_{l}^{\dagger j}b_{l}^{i}b_{k}^{j}+a_{k}^{\dagger i}b_{l}^{\dagger j}b_{l}^{i}a_{k}^{j})\\
 & = & \underset{klr}{\sum}(a_{s}^{\dagger r}b_{t}^{r}a_{k}^{\dagger i}a_{l}^{\dagger j}b_{l}^{i}b_{k}^{j}+b_{s}^{\dagger r}a_{t}^{r}a_{k}^{\dagger i}a_{l}^{\dagger j}b_{l}^{i}b_{k}^{j}+a_{s}^{\dagger r}b_{t}^{r}a_{k}^{\dagger i}b_{l}^{\dagger j}b_{l}^{i}a_{k}^{j}+b_{s}^{\dagger r}a_{t}^{r}a_{k}^{\dagger i}b_{l}^{\dagger j}b_{l}^{i}a_{k}^{j}).
\end{array}\label{eq:t1-1}
\end{equation}
}We explain the first term of equation (\ref{eq:t1}) $a_{k}^{\dagger i}a_{l}^{\dagger j}b_{l}^{i}b_{k}^{j}a_{s}^{\dagger r}b_{t}^{r}$
as an example. When $a_{k}^{\dagger i}a_{l}^{\dagger j}b_{l}^{i}b_{k}^{j}a_{s}^{\dagger r}b_{t}^{r}$
acts on the state of the many-body system, the system first annihilates
the $r$th particle in the state $t$. And then the $r$th particle
in the state $s$ is created. Then the system annihilates the $i$th
particle in the state $l$ after the annihilation of the $j$th particle
in the state $k$. After that the $j$th particle in the state $l$
and the $i$th particle in the state $k$ are created sequently. The
first term of equation (\ref{eq:t1-1}) describes the same process.
The actions of the operators of the first term of equation (\ref{eq:t1}) $a_{k}^{\dagger i}a_{l}^{\dagger j}b_{l}^{i}b_{k}^{j}a_{s}^{\dagger r}b_{t}^{r}$ on the states are
\begin{equation}
a_{k}^{i\dagger}a_{l}^{j\dagger}b_{l}^{i}b_{k}^{j}a_{s}^{r\dagger}b_{t}^{r}\left|\nu^{i}\nu^{j}\nu^{r}\right\rangle _{n}=\sqrt{\left\langle \nu\right\rangle _{nt}^{r}\left\langle \nu\right\rangle _{ns}^{r}\left\langle \nu\right\rangle _{nk}^{i}\left\langle \nu\right\rangle _{nl}^{i}\left\langle \nu\right\rangle _{nl}^{j}\left\langle \nu\right\rangle _{nk}^{i}}\left|\nu^{i}\nu^{j}\nu^{r}\right\rangle _{n}.
\end{equation}
Comparing with the action of the first term of equation (\ref{eq:t1-1}), we have
\begin{equation}
a_{s}^{r\dagger}b_{t}^{r}a_{k}^{i\dagger}a_{l}^{j\dagger}b_{l}^{i}b_{k}^{j}\left|\nu^{i}\nu^{j}\nu^{r}\right\rangle _{n}=\sqrt{\left\langle \nu\right\rangle _{ns}^{r}\left\langle \nu\right\rangle _{nt}^{r}\left\langle \nu\right\rangle _{nk}^{i}\left\langle \nu\right\rangle _{nl}^{j}\left\langle \nu\right\rangle _{nl}^{i}\left\langle \nu\right\rangle _{nk}^{j}}\left|\nu^{i}\nu^{j}\nu^{r}\right\rangle _{n}.
\end{equation}
They do describe the same process. So do other terms of these two equations.

It can be summarized as
\begin{equation}
\begin{array}{c}
\begin{array}{ccc}
i & j & r\end{array}\\
\left(\begin{array}{ccc}
l & k & t\\
k & l & s
\end{array}\right).
\end{array}
\end{equation}
So as to the following three terms of Eqs. (\ref{eq:t1}) and (\ref{eq:t1-1}).
We have $\left(\begin{array}{ccc}
l & k & t^{*}\\
k & l & s^{*}
\end{array}\right)$, $\left(\begin{array}{ccc}
l & k^{*} & t\\
k & l^{*} & s
\end{array}\right)$, and $\left(\begin{array}{ccc}
l & k^{*} & t^{*}\\
k & l^{*} & s^{*}
\end{array}\right)$. The superscript $*$ means the state in the conjugate space.

According to the basic assumption $a=b^{*}$and $[a,b]=0$, we have
$[a,a^{*}]=[b,b^{*}]=0$. In this case, the space of a Gentile many-body
system is the direct product of the Hilbert space and its complex
conjugate space. The space of Gentile many-body system can be expressed
as $S_{N}\otimes S_{N}^{*}\otimes U(m)\otimes U^{*}(m)$.

In the permutation group space $S_{N}\otimes S_{N}^{*}$, the generator
in Eq. (\ref{eq:a}) shows the exchanges in $S_{N}$ and $S_{N}^{*}$
respectively. These two exchanges read
\begin{equation}
\begin{array}{c}
ij\\
\left(\begin{array}{cc}
l & k\\
k & l
\end{array}\right)+
\end{array}\begin{array}{c}
ij\\
\left(\begin{array}{cc}
l & k^{*}\\
k & l^{*}
\end{array}\right).
\end{array}
\end{equation}
The irreducible representation of a single particle is written as
$\varepsilon_{m}\equiv\{1\}$ of $U(m)$ group. The creation of particle
goes to
\begin{equation}
\begin{array}{cc}
a^{\dagger}\left|0\right\rangle _{n}=\left|\varepsilon_{m},k\right\rangle \; & k=1\cdots m,\\
b^{\dagger}\left|0\right\rangle _{n}=\left|\varepsilon_{m}^{*},k\right\rangle \; & k=1\cdots m.
\end{array}
\end{equation}
The many-body Gentile system can be expressed as an $N$times single
space $\varepsilon_{m}^{N}=\varepsilon_{m}\otimes\cdots\otimes\varepsilon_{m}$
of $U^{N}(m)=U(m)\otimes\cdots\otimes U(m)$. Also we have 
\begin{equation}
a_{k_{1}}^{\dagger}\cdots a_{k_{N}}^{\dagger}\left|0\right\rangle _{n}=\left|\varepsilon_{m}^{N},k_{1}\cdots k_{N}\right\rangle ,1\leq k_{1}\cdots k_{N}\leq m.
\end{equation}
We label each irreducible representation by partition $\lambda(S_{N}\otimes S_{N}^{*})\otimes\lambda^{'}(U(m)\otimes U^{*}(m))$,
$(\lambda)=(\lambda_{1}\cdots\lambda_{m},\lambda_{1}^{*}\cdots\lambda_{m}^{*})$
and $\lambda=\lambda^{*}$ in the space $S_{N}\otimes S_{N}^{*}\otimes U(m)\otimes U^{*}(m)$,
where $\lambda_{1}+\cdots+\lambda_{m}=\lambda_{1}^{*}+\cdots+\lambda_{m}^{*}=N$
and $\lambda_{1}\geq\cdots\geq\lambda_{m}\geq0$, $\lambda_{1}^{*}\geq\cdots\geq\lambda_{m}^{*}\geq0$.
The space $S_{N}\otimes S_{N}^{*}\otimes U(m)\otimes U^{*}(m)$ can
be rewritten as
\begin{equation}
\left|\varepsilon_{m}^{N},k_{1}\cdots k_{N}\right\rangle =\sum\left|\varepsilon_{m}^{N},\lambda ii^{*},\lambda ll^{*}\right\rangle \left\langle \varepsilon_{m}^{N},\lambda ii^{*},\lambda ll^{*}\right|\left|\varepsilon_{m}^{N},k_{1}\cdots k_{N}\right\rangle ,
\end{equation}
where $\left|\varepsilon_{m}^{N},\lambda ii^{*},\lambda ll^{*}\right\rangle $
is called the Schur-Weyl basis \cite{s}, $\left|\lambda ii^{*}\right\rangle $
is the basis of the irreducible representation of $S_{N}\otimes S_{N}^{*}$,
and $\left|\lambda ll^{*}\right\rangle $ is the basis of the irreducible
representation of $U(m)\otimes U^{*}(m)$. When $\tau$ and $E$ act
on this state,
\begin{equation}
\tau\times E\left|\varepsilon_{m}^{N},\lambda ii^{*},\lambda ll^{*}\right\rangle =\lambda(\tau)_{ii^{*}}^{i^{'}i^{*'}}\lambda(E)_{ll^{*}}^{l^{'}l^{*'}}\left|\varepsilon_{m}^{N},\lambda i^{'}i^{*'},\lambda l^{'}l^{*'}\right\rangle ,
\end{equation}
where $\lambda(\tau)_{ii^{*}}^{i^{'}i^{*'}}$ and $\lambda(E)_{ll^{*}}^{l^{'}l^{*'}}$are
elements of the irreducible representation matrix $\lambda$.

\section{The Heisenberg model of many-body system in Gentile representation}

In this section, we give an example for demonstrating how to solve
an interacting many-body system using the group method in the occupation
number representation of Gentile statistics. The Heisenberg model
is a useful model to describe the interaction between spins. It usually
can be used to research the phase transition and the critical phenomenon
of magnetic systems and strongly correlated electronic systems. The
Hamiltonian is 
\begin{equation}
H=\underset{ij}{\sum}\frac{1}{2}\mathbf{S}_{i}\cdot\mathbf{S}_{j},
\end{equation}
where $\mathbf{S}$ is the spin and $\underset{ij}{\sum}$ indicate
the summation over all spin pairs. The model we discussed here is
a long-range interaction. The interactions are the same between any
pair of particles. There are two kinds of long-range interactions.
One likes the Coulomb interaction, the longer the distance, the weaker
the interaction is. The other likes oscillator potential, the longer
the distance, the stronger the interaction is. The interaction we
discussed is the long-range interaction between the Coulomb interaction
and the oscillator interaction.

It is easy to check that this Hamiltonian can be rewritten as(for
example, a system only has two spins)
\begin{equation}
H=\underset{ij}{\sum}\tau_{ij}+constant=P(2,1^{\nu-2})+constant,
\end{equation}
and the constant can be ignored.

According to the theorem in section $2$, Eq. (\ref{eq:=000026})
shows the relation between the Casimir operator of the unitary group
and the conjugacy-class operator of the permutation group in the Gentile
statistical representation. We substitute the Casimir operator of
the unitary group for the conjugacy-class operator of the permutation
group.
\begin{equation}
H=\cos^{-1}\frac{2\pi}{n+1}(\frac{1}{2}C_{2}-C_{1}-2\underset{ki}{\sum}J(N_{k}^{i})).
\end{equation}
Here, we have $m=2.$ This relation tells us that the energy spectrum
of the many-body intermediate statistical system is related to the
maximum occupation number $n$ and the particle number of the system.
In the Gentile statistical representation, it is easy to calculate
the energy spectrum of a many-body system. Ref. \cite{c} gives the
eigenvalues of $C_{1}$and $C_{2}$ as $\left\langle C_{1}\right\rangle =S_{1}$and
$\left\langle C_{2}\right\rangle =S_{2}-(m-1)S_{1}$, where 
\[
S_{l}=\sum_{i=1}^{m}\left[(a_{i}+m-i)^{l}-(m-i)^{l}\right],
\]
and $a_{i}$ is the partition of the irreducible representation of
$U(m)$ \cite{c,c1,z,z1,z2}.

When $n\rightarrow\infty$ and $n=1$, it returns the Bose system
and Fermi system respectively: 
\begin{equation}
\begin{array}{c}
H=\frac{1}{2}C_{2}-C_{1},\qquad\textrm{Bose case}\\
H=-(\frac{1}{2}C_{2}-C_{1}).\qquad\textrm{Fermi case}
\end{array}
\end{equation}
The exact value of energy spectrum and degeneracy are discussed in
Ref. \cite{G}.

\section{Discussion and conclusion}

The exactly solvable models are very important in physics, but they
are rare. Various approximation methods are introduced to solve the
problems such as mean-field approximation and the Jordan-Wigner transformation.
Group is a useful method in dealing with many-body interacting systems
in physics. As is known, all finite groups are isomorphic to the permutation
groups, and all infinite group are isomorphic to the unitary groups.
It makes sense to construct the relation between the permutation group
and the unitary group. Gentile statistics whose maximum occupation
number is finite has the general form of fractional statistics. Bose
and Fermi statistics are two limits of Gentile statistics. The occupation
number representation of Gentile statistics makes it easier to solve
the many-body interacting systems, because the state is just represented
by the occupation number. We reveal the relation between the Casimir
operator of the unitary group and the conjugacy-class operator of
the permutation group in Gentile statistics representation. The relation
recovers Bose and Fermi statistics when the maximum occupation number
is infinity and one. The Heisenberg model is an important interacting
spin model in statistical physics. On the basis of the relation between
the permutation group and the unitary group, we calculate the Heisenberg
model by using the Casimir operator of the unitary group $U(m)$ and
give a series of exactly solvable models as examples. The Heisenberg
model with interactions between each pair of particles is the special
case at $m=2$. 

In this paper, we introduced a group method to construct the exactly
solvable model. We mapped an interacting spin system onto an equivalent
intermediate statistical system (Gentile representation). In other
words, we constructed the correlation between interacting spin system
and intermediate statistical system. This is an intermediate statistical
description of interacting spin system. Our method here does not mean
that the intermediate statistical system obeys Gentile statistics.
What we do is dealing with the many-body interacting system in the
Gentile representation. And this is the best way to understand some
physical systems which we are not familiar with. For example, quantum
interacting spin systems are standard techniques on quantum computation
and quantum information. Our method provides a new approach to simulate
the quantum computer using intermediate statistical system. At the
same time, this method makes it possible to deal with intermediate
statistical system by quantum computer.

\section{Acknowledgments}
The research was supported by the Fundamental Research Funds For the
Central Universities No.2020JKF306.

%\appendix
%\section{Some title}
%Please always give a title also for appendices.

%\acknowledgments%崑仍
%%%%%%%%%%屎猟潤崩

%\begin{thebibliography}{99}

%\end{thebibliography}\endgroup

%\bibitem{a}
%Author, \emph{Title}, \emph{J. Abbrev.} {\bf vol} (year) pg.

%\bibitem{b}
%Author, \emph{Title},
%arxiv:1234.5678.

%\bibitem{c}
%Author, \emph{Title},
%Publisher (year).

% Please avoid comments such as "For a review'', "For some examples",
% "and references therein" or move them in the text. In general,
% please leave only references in the bibliography and move all
% accessory text in footnotes.

% Also, please have only one work for each \bibitem.

%\end{thebibliography}

\end{document}